\documentclass[aps,prl,twocolumn,superscriptaddress,showpacs]{revtex4}
\usepackage{graphicx}
\usepackage{amsmath}
\usepackage{natbib}
\begin{document}
\title{Vortex interaction, chaos and quantum probabilities}
\author{D. A. Wisniacki}
\affiliation{Departamento de F\'{\i}sica ``J. J. Giambiagi'', FCEN, UBA,
 Pabell\'on 1, Ciudad Universitaria,
 1428 Buenos Aires, Argentina.}
\author{E. R. Pujals}
\affiliation{IMPA--OS, Dona Castorina 110,
22460--320, Rio de Janeiro, Brasil.}
\author{F. Borondo}
\email[E--mail address: ]{f.borondo@uam.es}
\affiliation{Departamento de Qu\'{\i}mica C--IX,
 Universidad Aut\'onoma de Madrid,
 Cantoblanco--28049 Madrid, Spain.}
\date{\today}
\begin{abstract}
The motion of a single vortex is able to originate chaos in the quantum
trajectories defined in Bohm's interpretation of quantum mechanics.
In this Letter, we show that this is also the case in the general situation,
in which many interacting vortices exist.
This result gives support to recent attempts in which Born's probability
rule is derived in terms of an irreversible time evolution to equilibrium,
rather than being postulated.
\end{abstract}
\maketitle
%
Despite the impressive success of quantum mechanics along the past century,
Born's probability rule: $\rho=|\psi|^2$,
one basic cornerstone in its standard formulation,
still remains a postulate.

Due to its relevance, this fundamental issue have been revisited
in the last years \cite{Zurek,Wallace,Valentini}, in an effort to
make probability an emergent phenomenon \cite{Adler}.
Notice, for example, that basic ingredients in the decoherence programme,
such as reduced density matrix, are based on Born's rule.
In this respect, Zurek \cite{Zurek} introduced environment assisted
invariance (``envariance´´), a causality related symmetry of
quantum entangled systems, to derive Born's rule.
Wallace and Deutsch \cite{Wallace} reported another approach
based on classical decision theory in the context of Everett many
worlds interpretation.
Another interesting point of view is that of Valentini and Westman
\cite{Valentini}, who argued, using the (also causal)
de Broglie--Bohm (dBB) \cite{Bohm} quantum formalism \cite{footnote},
that probabilities have a dynamical origin, holding a status similar
to that of thermal probabilities in ordinary statistical mechanics.
Indeed, the standard distribution is obtained as the time evolution
towards the equilibrium of initial non--equilibrium states,
$\rho \neq |\psi|^2$, this taking place with a (exponential)
decrease in the associated coarse--grained $H$--function.
Underlying to this argument is the assumption that there is an
effective chaotic dynamics in the dBB trajectories,
something that should not be taken for granted.
Unfortunately, most published results along this line are
rather inconclusive.
However, very recently singularities in the wave function
giving rise to vortices,
have been proven to play a prominent role in the problem.
In Ref.~\onlinecite{Pujals}, the case of a single vortex was considered,
arriving at the conclusion that the motion of an isolated vortex
is enough to originate chaos in Bohmian trajectories.
However, nothing is known about the picture emerging in the general
situation, in which many vortices exist.
For this case, in addition to the vortex dynamics,
a great deal of interaction should be expected,
for example through a mechanism of creation/annihilation of pairs
with opposite circulations.
In this respect, Frisk \cite{Frisk} conjectured the importance of the
number of nodes inducing mixing behavior in dBB trajectories.
In view of the paramount importance of the problem under discussion,
a deep understanding of this issue, similar to that in statistical
mechanics, is highly desirable.

In this Letter, we address the question of the complexity of Bohmian
trajectories by presenting a systematic
numerical study in which we show that the chaotic regime,
due to the dynamics and interactions of the existing vortices,
is the general rule in dBB trajectories even in the absence of
nonlinear terms in the physical potential.
This gives rise to a noticeable complexity that can be quantified
in terms of common indicators, such as Lyapunov exponents.
As an added bonus, the use of such indicator allow to explore some
differences existing with the classical case.

In the dBB theory \cite{Bohm} the state of the system is described
by a pilot wave function, customarily expressed in polar form,
$\psi(\mathbf{r},t) = R(\mathbf{r},t) \; {\rm e}^{iS(\mathbf{r},t)}$
($\hbar=1$ throughout the paper), and the position of the particles,
$\mathbf{r}$ \cite{Durr}.
The dynamical evolution of this two quantities is given by the
time--dependent Schr\"odinger equation and the guidance equation,
respectively:
%
\begin{equation}
  \mathbf{v} = \dot{\mathbf{r}} = \frac{\nabla S}{m} = \frac{\rm i}{2m}
      \frac{\psi\nabla\psi^*- \psi^*\nabla\psi}{|\psi|^2},
  \label{eq:1}
\end{equation}
where $m$ is the mass of the particle.
Quantum trajectories, which make of dBB a true theory of quantum motion
\cite{Holland}, can be obtained by numerical integration of this equation.
The velocity field (\ref{eq:1}), guiding these trajectories,
presents singularities giving rise to vortices in the associated
probability fluid.
This occur only at points where the wave functions vanishes
(isolated points in 2--dof systems, lines in 3--dof systems, etc.)
and the phase $S$ is singular.
Moreover, and due to the single--valuedness of the wave function,
the circulation, $\Gamma$, around a circuit, ${\cal C}$, encircling
a vortex must be quantized \cite{Dirac,Birula1} according to
%
\begin{equation}
  \Gamma = m \oint_{\cal C} \; \dot{\mathbf{r}} \cdot d \mathbf{r} =
    \oint_{\cal C} \; \nabla S \cdot d \mathbf{r} = 2 \pi n,
\end{equation}
where $n$ is a integer.
This implies that the velocity must diverge at the vortex
\cite{Birula2,Falsaperla}.

The aim of our work is to study the behavior of the quantum
trajectories of a system in the general situation in which a large
number of interacting vortices exists.
One of the simplest systems where this problem can be set up is the
2--dof rectangular billiard whose classical dynamics are integrable.
In this way, any observed complexity is only due to quantum effects,
without any contribution of chaos coming from forces derived
a physical potential.
The dimensions of the rectangle are taken so that the smallest side is
equal to 1, and the total area amounts to $A=1+\pi/2$,
so that the corresponding eigenfunctions are
$\phi_{n_x,n_y}(x,y)=(2/A^{1/2})\sin(n_x \pi x)\sin(n_y \pi y/A)$,
with wave numbers $k=(2mE)^{1/2}=\pi(n_x^2+n_y^2/A^2)^{1/2}$.
The initial pilot wave is constructed as a linear combination,
$\psi(x,y)=\sum_{n_x,n_y}^N c_{n_x,n_y} \phi_{n_x,n_y}(x,y)$,
of $N$ such eigenfunctions with random coefficients.
We will systematically vary this function by considering increasing
values of the wave number mean value, $\langle k \rangle$, and $N$.
For this purpose, the eigenstates entering into the linear combination,
giving initially the pilot wave function, are selected in the following way.
At any given value of the energy, a ``central'' state is determined by
choosing the nearest integers, $n_x=n_y$, fulfilling the energy condition;
the rest of intervening states are then chosen as the $N-1$ states which
are closest in energy to the central one.

\begin{figure}[t]
 \includegraphics[width=7.0cm]{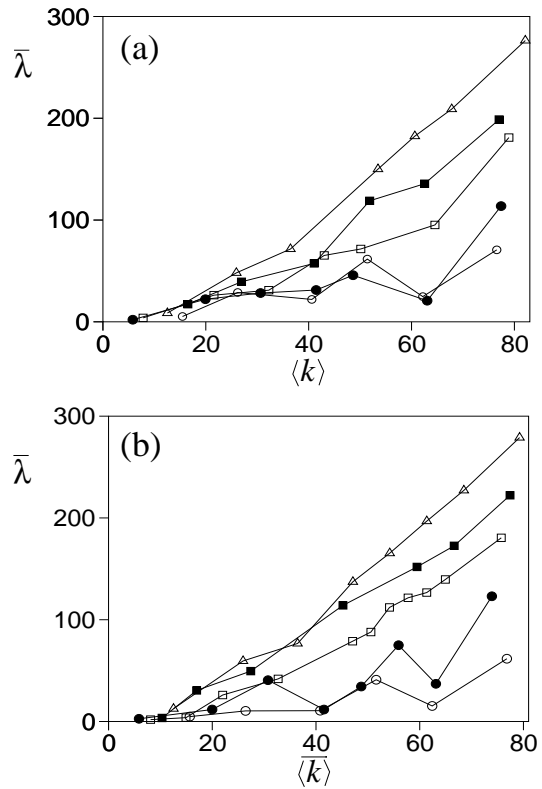}
   \caption{Averaged Lyapunov exponent as a function of the mean wavenumber
   for quantum trajectories on a rectangular billiard of smallest side
   length equal to 1 and area $1+\pi/2$.
   The initial pilot wave function is taken as a linear combination
   of $N$ rectangle eigenfunctions, being
   $N$=3 (open circles),
       5 (full circles),
      11 (open squares),
      20 (full squares),
      30 (open triangles).
   In (a) the average is carried out over 50 randomly selected initial
   conditions, while in (b) the coefficients of the initial pilot wave
   function were also varied.}
  \label{fig:1}
\end{figure}

To gauge the complexity of our system we take, similarly to what it
is well established in the usual chaos theory,
the Lyapunov exponent of the quantum trajectories, $\overline{\lambda}$,
statistically averaged over many initial random conditions.
The corresponding results for $\overline{\lambda}$, as a function of
$\langle k \rangle$ for different values of $N$,
are shown in Fig.~\ref{fig:1}.
%
As can be seen, $\overline{\lambda}$ grows systematically both with
$\langle k \rangle$ and $N$, showing a mean tendency which is
approximately linear after a threshold at
$\langle k \rangle_{\rm th} \sim 30$.
To check that this conclusion is not an artifact of the way in
which the initial wave function (i.e.\ the coefficients
entering in the linear combination) has been chosen,
we have repeated the same calculation using a different averaging
procedure for the Lyapunov function.
In this second calculation we change not only the initial position
of each quantum trajectory in the averaging ensemble,
but we also change randomly the coefficients in the corresponding
initial pilot wave.
This double average defines a new mean wave number that will be
denoted by $\overline{\langle k \rangle}$.
The results are shown in Fig.~\ref{fig:1}(b),
where it is seen that they follow a behavior totally equivalent to that
obtained in the previous case [curves in the part (a) of the figure].
This fact indicates that our conclusion is robust.
Making it quantitative, the final increasing linear tendency of
the mean Lyapunov exponent is given by the expression:
$\overline{\lambda}=0.15N\langle k \rangle$.

Following Frisk \cite{Frisk} let us try now to explain these results
in terms of the number of vortices existing in the system.
This is a sensible assumption, since it is at these points where the
complexity in the quantum trajectories is originated \cite{Pujals}.
This is, however, not a straightforward task,
since the number of vortices associated to each pilot wave function
varies with time, significantly fluctuating around the mean value,
by creation and annihilation of pairs of vortices with opposite
circulations.
Taking this into account, we compute the time average of the
number of vortices (with circulation in a given sense),
${\overline N}_v$, as a numerical indicator characterizing
the collective effect of the vortices.

%
\begin{figure}[t]
 \includegraphics[width=7.0cm]{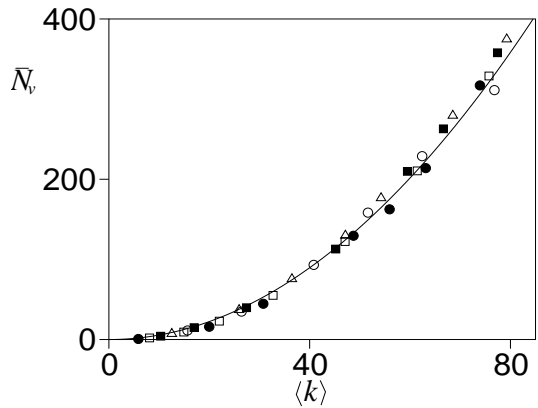}
   \caption{Time average of the number of vortices with
   a given sense of the circulation for the  different cases
   shown in Fig.~\protect\ref{fig:1}(a).
   Full line corresponds to the analytical expression
   $N_v^{\rm max}=(A/4\pi^2)k^2-[(1+A)/2\pi]k$,
   giving a rough estimate of the maximum number of
   vortices fitting in our billiard (see text for details).}
  \label{fig:2}
\end{figure}
The corresponding results, obtained for the same conditions
of Fig.~\ref{fig:1}(a), are shown in Fig.~\ref{fig:2}.
As can be seen the mean number of vortices grows quadratically
with $\langle k \rangle$, but it is completely independent of $N$,
being the same at a given energy, regardless of the number of
eigenfunctions contributing to the pilot wave function used in
the computation of the quantum trajectories.
This numerical calculation clearly indicates, contrary to Frisk
expectations, that the number of vortices alone is not enough to
explain complexity found generically in Bohmian trajectories.
Furthermore, the result in Fig.~\ref{fig:2} can be understood
by considering the following rough estimate of the maximum
possible number of vortices that can fill our billiard.
Assuming that the minimum area in configuration space required
for the existence of a vortex is given by the magnitude of the
squared de Broglie wavelenght,
$\lambda_{\rm DB}^2=4\pi^2/\langle k \rangle^2$,
and taking into account the boundary effects at the walls,
the maximum number of vortices in the billiard should be given by
$N_v^{\rm max} = (A/4\pi^2)\langle k\rangle^2
-[(1+A)/2\pi]\langle k \rangle$,
expression that agrees perfectly well with the computed data,
as shown by the full line in Fig.~\ref{fig:2}.

Since ${\overline N}_v$ is not enough to explain the behavior
of $\overline{\lambda}$ in Fig.~\ref{fig:1}(a),
let us try now the second momentum of the corresponding temporal
distribution.
Notice that this is equivalent to assume that the origin of the
complexity of quantum trajectories is, in the general case,
the interaction responsible for the vortex creation/annihilation
mechanism.
%
\begin{figure}[t]
 \includegraphics[width=7.0cm]{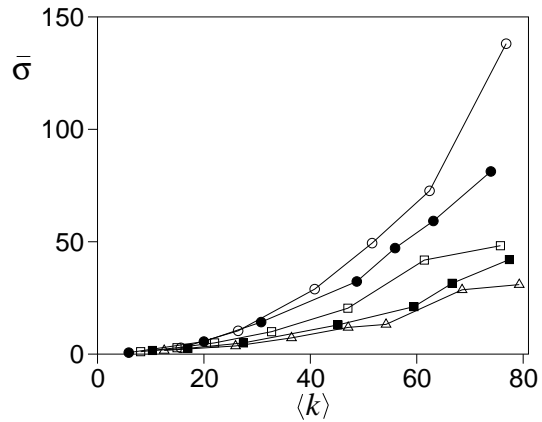}
   \caption{Time average of the mean root deviation on the
   number of vortices with a given sense of the circulation
   for the different cases shown in Fig.~\protect\ref{fig:1}(a).}
  \label{fig:3}
\end{figure}
The results are plotted in Fig.~\ref{fig:3},
where it can been seen the time average of the mean root deviation
on the number of vortices,
$\overline{\sigma}$, grows monotonically with ${\overline k}$
and decreases with $N$, i.e.\ the more complicated the function
is the smallest dispersion is found.
Moreover, this scaling dependence can be formulated in quantitative
terms as: ${\overline \sigma}=0.035 \langle k \rangle^2 N^{-1/2}$.
The quadratical dependence on $k$ is obvious, since it is a
consequence of the functional form exhibited by ${\overline N}_v$,
and accordingly the mean root deviation can be also expressed as:
${\overline \sigma}=0.54 {\overline N}_v N^{-1/2}$.
What it is interesting, is that ${\overline \sigma}$ contains an
additional inverse dependence on $N$.
This result, if examined carefully (as will be discussed below),
is in agreement with the systematic behavior previously found for
the quantum trajectories Lyapunov exponent in Fig.~\ref{fig:1}(a).
Finally, by putting these two components together,
the complexity of quantum trajectories as quantified by
the averaged Lyapunov parameter can be solely expressed in
terms of vortex properties, in the following way:
${\overline\lambda}=0.17{\overline N}_v^{5/2}{\overline \sigma}^2$.

\begin{figure}[t]
 \includegraphics[width=7.0cm]{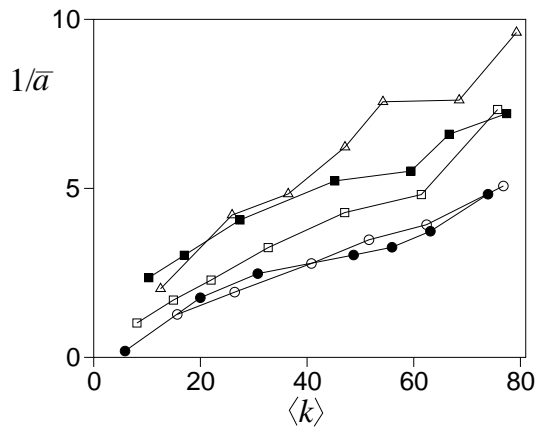}
   \caption{Area of the largest compact vortex free region in
   the rectangular billiard corresponding to the different cases
   shown in Fig.~\protect\ref{fig:1}(a).}
  \label{fig:4}
\end{figure}

Now, let us rationalize the behavior found for ${\overline \sigma}$
as a function of $N$.
Fluctuations in the number of vortices, as a result of pair
annihilations, leave areas of the billiard depopulated from them.
The existence of these areas decreases the value of the mean
Lyapunov exponent, ${\overline \lambda}$, since,
according to the results in Ref.~\onlinecite{Pujals},
in these areas the complexity in the dynamics of the quantum
trajectories is smaller.
The corresponding flux of the quantum fluid is more laminar
in those regions, only getting turbulent close to the
remaining vortices.
The argument can be made quantitative by performing the
following calculation.
At the same times used to compute the averages presented
in Fig.~\ref{fig:1}(a) and with the same wave functions we
calculate the position of the corresponding vortices.
We then put a fine grid in configuration space and mark
differently the squares with and without at least a vortex
inside them.
From this matrix we compute, at each value of $\langle k \rangle$,
the size of the largest compact region without any vortex, $a$.
Finally, the areas of these vortex--free regions are averaged
along trajectories and with respect to initial positions.
Now, we take the inverse of this quantity to get a magnitude
with the same dependence on the complexity as ${\overline \lambda}$.
The results are shown in Fig.~\ref{fig:4}.
As can be seen, $1/{\overline a}$ increases linearly with
$\langle k \rangle$ for each value of $N$, and also it keep the
same dependence with $N$ as found in Fig.~\ref{fig:1}(a) for
${\overline \lambda}$.
In this way, our calculations numerically proof that
there are two mechanisms originating chaos in this problem.
The first one has a local character and corresponds
to the randomization effect due to moving
vortices on the quantum trajectories \cite{Pujals},
while the second global one is the appearance of vortex--free
areas due to annihilation of pairs with opposite circulations.
%

In order to close this discussion, it should be remarked that
other calculations have been performed in order to rule out the
possibility that others effects are relevant in the behavior of
the complexity of quantum trajectories.
In particular, the most obvious ones are the kinematical magnitudes.
The most straightforward check is to study the velocity
and acceleration distributions of the vortices.
They can be calculated from the expression
%
\begin{equation}
  \mathbf{v} = \frac{1}{2m{\rm i}} \;
  \frac{\mathbf{r}\times(\mathbf{w}\times\mathbf{w}^*)}
    {|\mathbf{w}\cdot\mathbf{r}|^2},
\end{equation}
where $\mathbf{w}\equiv\nabla\psi[\mathbf{r}(t)]$ \cite{Birula2}.
Our results clearly indicate that neither of them shows any obvious
systematic, as it happens with the correlation found by us between
the Lyapunov exponent and the number of vortices.

Finally, another interesting point to discuss here is the classical
limit of our results.
The results in Fig.~\ref{fig:1} indicate that the averaged Lyapunov
exponent do not seem to vanish as
$\langle k \rangle \rightarrow \infty$.
This means that in this semiclassical limit, the effect of vortices
inducing chaos and complexity in the quantum trajectories,
do not disappear.
The behavior is not, however, unexpected since it has also been
found in similar problems.
For example, in Ref.~\cite{Sanz2} it was shown that the non--local
character of the Bohmian trajectories (avoiding crossing, for example)
survives when the quantum effects, represented by the quantum potential $Q$,
were made to disappear by increasing the mass of the incident particle
in a realistic model for rainbow diffraction in atom--surface scattering.
The quantum trajectories simply mimicked the classical distributions
without ever reaching strictly the classical limit.
Also, Bowman \cite{Bowman} pointed out that the Bohmian classical limit
can only be achieved by combination of narrow packets, mixing states,
and environment decoherence.
Certainly, further calculations, considering much higher values of
$\langle k \rangle$, are needed in order to fully confirm
our results and clarify this issue.

Summarizing, in this Letter we have numerically shown that chaos is the
generic scenario for quantum trajectories, in the situation
in which many interacting vortices exists.
In this way the picture started with Ref.~\onlinecite{Pujals},
in which the effect of only one vortex was studied, is completed.
Moreover, we have quantified this assertion with the aid of the
Lyapunov exponent as a numerical indicator of complexity.
Our results show that the behavior of this quantity
depends on the number of vortices of the pilot wave function,
and only the first two momenta of the corresponding distribution
are required to explain it satisfactorily.
This result is interesting, since it shows the interplay between the
quantum phase, which appears in the guiding equation (\ref{eq:1}),
and quantum probabilities, these being the two components in which
the wave function is separated in the  Bohmian formulation of
quantum mechanics.
\begin{acknowledgments}

This work was supported by CONICET and UBACYT (X248) (Argentina),
and MCyT (Spain) under contract BQU2003--8212.
\end{acknowledgments}
%

%
\end{document}